# Data-driven Sensor Deployment for Spatiotemporal Field Reconstruction

Jiahong Chen[1]

*(1. Department of Mechanical Engineering, The University of British Columbia, Vancouver, BC, Canada)*

**Abstract:** This paper concerns the data-driven sensor deployment problem in large spatiotemporal fields. Traditionally, sensor deployment strategies have been heavily dependent on model-based planning approaches. However, model-based approaches do not typically maximize the information gain in the field, which tend to generate less effective sampling locations and lead to high reconstruction error. In the present paper, a data-driven approach is developed to overcome the drawbacks of the model-based approach and improve the spatiotemporal field reconstruction accuracy. The proposed method can select the most informative sampling locations to represent the entire spatiotemporal field. To this end, the proposed method decomposes the spatiotemporal field using principal component analysis (PCA) and finds the top $r$ essential entities of the principal basis. The corresponding sampling locations of the selected entities are regarded as the sensor deployment locations. The observations collected at the selected sensor deployment locations can then be used to reconstruct the spatiotemporal field, accurately. Results are demonstrated using a National Oceanic and Atmospheric Administration sea surface temperature dataset. In the present study, the proposed method achieved the lowest reconstruction error among all methods.

## 1 Introduction

Recently, researchers have endeavored to monitor and reconstruct an extensive spatiotemporal field using limited resources. Both model-based and data-driven sensor deployment strategies have been investigated to obtain near-optimal sparse sampling locations that can retrieve maximal information from the spatiotemporal field. Especially, data-driven approaches, such as sparse learning, investigate linear mapping between low-dimensional sparse observations and the original high-dimensional signal. Therefore, sparse learning approaches, such as compressive sensing and principal component analysis, have been widely used to recover signals from limited observations collected in the near-optimal sampling locations [1]–[3].

Typically, model-based approaches have utilized robotic sensors to establish environmental models and find suitable sensor deployment locations. In these approaches, a fleet of robotic sensors is sent out to explore a field of interest for a generalized understanding of the environment. Near-optimal sampling locations can then be computed based on the environmental model. Nguyen and colleagues developed a strategy for driving robotic sensors in a mobile wireless network to efficiently monitor the environment and predict spatial phenomena [4]. Notably, they considered the case where the received sensing locations can be inaccurate. Their proposed method can design optimal sampling paths and locations for mobile robotic sensors, given the localization uncertainties. Besides, energy constraints are considered, which makes this strategy practical for mobile sensor networks. Subsequently, Ma and his colleagues developed a learning and planning method for robotic sensors to sample the environment [5], [6]. They proposed a framework that can persistently monitor the environment of interest by learning and updating the attributes of the spatiotemporal environmental model. Besides, Dunbabin et al. addressed the existing sampling limitations by fusing the sampled data from two robotic sensor systems [7]. The developed system can obtain both continuous spatial coverage and temporal measurements across an entire water body.

Furthermore, researchers have focused on finding sampling locations that can maximize the information gain from a field while generating an accurate environmental model. Kashino and colleagues developed an optimal sensor deployment strategy for mobile target detection in a spatiotemporal field [8]. They considered the case that the environment of interest was greater than the size of the detectable region of all sensors, which is referred to as sparse coverage. The proposed optimal sensor deployment strategy efficiently and precisely determines the movement of the robotic sensors according to the highest possible likelihood. Quann et al. proposed an energy-efficient sampling strategy by considering both spatiotemporal field uncertainty minimization and energy consumption of the robotic sensor [9]. Hitz et al. proposed a novel evolutionary strategy for planning informative paths, which achieved rich information from a field while obeying traveling path budget constraints [10]. They also included an adaptive replanning method to update paths to retrieve a lower field estimation uncertainty.

In contrast to model-based approaches, data-driven approaches do not build environmental models. Instead, they optimize the sampling locations directly using historical sampled data. For example, sparse learning methods, such as compressive sensing and principal component analysis (PCA), can project a high-dimensional spatiotemporal field onto a low-dimensional space [23, 25]. The low-dimensional space can then be used to reconstruct the whole spatiotemporal field with high accuracy. Brunton et al. proposed a compressive sensing-based sensor placement method for high-dimensional system classification [24]. Later, Manohar et al. proposed an optimal sensor deployment algorithm based on QR factorization and PCA. They compared different types of data-driven reconstruction methods and found that PCA-based methods can outperform compressive sensing-based methods [3]. Hence, data-driven methods based on PCA can be utilized to learn a projection basis from the training data. Singular value decomposition (SVD) can then be used to tailor unnecessary information and retain least sampling locations. The entire spatiotemporal field can be subsequently reconstructed based on the observations.

In the present paper, a data-driven approach is proposed for sensor node deployment and spatiotemporal field reconstruction. The proposed approach deploys sensor nodes that represent the entire spatiotemporal field near-optimally given a set of historical sampled data. Observations collected in the deployed locations can then be used to reconstruct the spatiotemporal field by learning the sparse representation mapping using a data-driven approach.

# 2 PRELIMINARIES

## 2.1 Model-based Sampling Optimization

Traditionally, the sampling location optimization is based on an environmental model that is learned from the training data [11, 12, 26]. The environment can be modeled as:

$$v_\psi[t] = F(\mathbf{A}, \mathbf{S}_\psi, \psi, t),$$

where $\psi$ is an arbitrary point in the two-dimensional field $\mathcal{A}$, matrix $\mathbf{A}$ records temporal information, matrix $\mathbf{S}_\psi$ records spatial information, and $t$ is the timestamp.

In this manner, the environmental model can be represented by spatial information and temporal information, separately [12]. The spatial information can be represented by $\mathbf{S}_\psi$ and the time-varying coefficient vector $\mathbf{x}[t]$ [13, 14]:

$$v_\psi[t] = \mathbf{S}_\psi \mathbf{x}[t] + v[t], \quad (2.1.1)$$

where $v[t] \sim N(0, \mathbf{R}_v)$ is Gaussian white noise, and $\mathbf{R}_v$ is its covariance matrix. The temporal information can be represented by $\mathbf{A}$ and the time-varying coefficient vector $\mathbf{x}$:

$$\mathbf{x}[t+1] = \mathbf{A}\mathbf{x}[t] + w[t], \quad (2.1.2)$$

where $w[t] \sim N(0, \mathbf{R}_w)$ is Gaussian white noise and $\mathbf{R}_w$ is the covariance matrix. Note that matrix $\mathbf{A}$ and matrix $\mathbf{R}_w$ can be obtained from sampling data [15,16].

Then, the effectiveness of the sampling locations can be evaluated using control methods. For example, Kalman filter is a suitable control approach for estimating time-varying coefficients in state-space models. The prior error covariance matrix of the state-space model can be computed by the Kalman filter:

$$\mathbf{P}[t+1|t] = \mathbf{A}\mathbf{P}[t|t]\mathbf{A}^T + \mathbf{R}_w$$

where:

$$\mathbf{P}[t|t] = \mathbf{P}[t|t-1] - \mathbf{P}[t|t-1]\mathbf{S}_{\psi_i}^T \times$$
$$\left(\mathbf{S}_{\psi_i}\mathbf{P}[t|t-1]\mathbf{S}_{\psi_i}^T + \mathbf{R}_v\right)^{-1}\mathbf{S}_{\psi_i}\mathbf{P}[t|t-1] \quad (2.1.3)$$

Then, the effectiveness of the sampling locations can be evaluated by the maximum eigenvalue of the prior covariance matrix [17]. Hence, the model-based informative sampling strategies seek the deployment strategy that has minimum largest eigenvalue:

$$\Gamma(\mathbf{P}^\mathcal{N}) = \max_{n \in \mathcal{N}} Eigen(\mathbf{P}^n), \quad (2.1.4)$$

where $\mathcal{N}$ is the deployment strategy containing the sensor deployment locations, $\mathbf{P}^n$ indicates the prior error covariance for sampling location $n \in \mathcal{N}$.

## 2.2 Data-driven Sampling Optimization

In contrast to model-based approaches, data-driven approaches optimize sampling locations directly from training data, and not based on models. PCA is optimal for recovering high-dimensional signals from unknown contents. It converts correlated observations into principal components that have hardly any linear redundancy. The training data $\mathbf{\Phi}_{\text{train}}$ can be represented as:

$$\mathbf{\Phi}_{\text{train}} = \mathbf{\Psi} \cdot \mathbf{S} \cdot \mathbf{V}, \quad (2.2.1)$$

where $\mathbf{\Psi}$ is the left singular vectors of $\mathbf{\Phi}_{\text{train}}$, $\mathbf{S}$ is a rectangular diagonal matrix of singular values of $\mathbf{\Phi}_{\text{train}}$, and $\mathbf{V}$ is the right singular vectors of $\mathbf{\Phi}_{\text{train}}$.

Then, the original signal can be represented only using first $r$ largest eigenvalues and the singular vectors:
$$\mathbf{\Phi}_{\text{train}} \approx \mathbf{\Phi}^*_{\text{train}} = \mathbf{\Psi}_r \cdot \mathbf{S}_r \cdot \mathbf{V}_r \quad (2.2.2)$$

In this manner, high-dimensional signals can be represented by a few sparse representations in the field. Then, sparse learning approaches can be used to study the relationship between the compressed signal and the original signal.

Denote the spatiotemporal field as a high dimensional state $\phi \in \mathbb{R}^m$, and the spatiotemporal dynamics of $\phi$ can be captured by the low dimensional observations. The selected observations ($\mathbf{y}$) can be represented by a measurement matrix $\mathbf{C} \in \mathbb{R}^{r \times m}$, where $r$ is the number of measurements:
$$\mathbf{y} = \mathbf{C} \cdot \phi. \quad (2.2.3)$$

Note that $\phi$ can be sparsely represented by a principal basis:
$$\phi = \mathbf{\Psi}_r \cdot \mathbf{a} \quad (2.2.4)$$
where $\mathbf{\Psi}_r \in \mathbb{R}^{m \times r}$ is the principal basis learned from training data, and $\mathbf{a} \in \mathbb{R}^r$.

**2.3 Problem Formulation**

Mathematically, the sparse signal reconstruction can be expressed as the projection from a low-dimensional observation space to a high-dimensional signal:
$$\widetilde{\phi} \xleftarrow{\xi} \mathbf{y} \colon \mathbb{R}^m \leftarrow \mathbb{R}^r.$$

As $r \ll m$, this system is underdetermined. Hence, there are infinite solutions. Hence, the sampling locations can be optimized, and the mean error may utilized to evaluate the reconstructed spatiotemporal field:
$$\arg\min_{\xi, \mathbf{y}} ||\phi - \xi(\mathbf{y})||$$

Key variables used in this paper are defined in Table 1:

Table 1: Definition of key variables.

| Variable | Definition |
|---|---|
| $\mathcal{A}$ | a two-dimensional environment of interest |
| $\mathbf{A}$ | temporal information matrix |
| $\mathbf{C}$ | measurement matrix |
| $\mathcal{N}$ | deployment strategy |
| $\mathbf{P}$ | estimation error matrix |
| $\psi$ | a location in the two-dimensional environment |
| $\phi$ | high-dimensional spatiotemporal field |
| $\mathbf{\Phi}$ | time series training data sampled from spatiotemporal field |
| $\mathbf{\Psi}$ | principal basis |
| $\mathbf{R}$ | covariance matrix of Gaussian white noise |
| $\mathbf{S}$ | rectangular diagonal matrix |
| $\mathbf{S}_\psi$ | the spatial information matrix at location $\psi$ |
| $\mathbf{V}$ | right singular vectors |
| $\mathbf{x}[t]$ | time-varying coefficient vector |
| $\mathbf{y}$ | observations in the field |

# 3 Proposed Method

## 3.1 Sparse Learning for Spatiotemporal Reconstruction

According to Equation (2.2.2), high-dimensional states $\phi \in \mathbb{R}^m$ can be represented as linear combinations of $r$ orthogonal eigenmodes $\psi$. This low-dimensional state can be learned from training data by pruning the SVD basis. In this manner, $\phi$ may be decomposed to produce temporal and spatial coefficients [3], and the linear combination of $\phi$ shown in Equation 2.2.4 can be represented as:
$$\phi_i = \sum_{k=1}^{r} a_k(t_i) \psi_k(\phi), \quad (3.1.1)$$
where $\phi_i$ is the spatiotemporal field at time $i$ and $\psi_k(\phi)$ is the time independent spatial coefficient. The temporal coefficient $a_k(t_i)$ varies with time $t_i$.

Next, $\psi_k$ and $a_k$ can be computed by SVD. Training data $\mathbf{\Phi} = [\phi_1 \phi_2 \cdots \phi_M]$ are given for $M$ snapshots. The tailored SVD basis consists of orthonormal left singular vectors $\mathbf{\Psi}$, right singular vectors $\mathbf{V}$ and the diagonal matrix $\mathbf{S}$. Hence, the training data can be represented as:
$$\mathbf{\Phi} = \mathbf{\Psi} \cdot \mathbf{S} \cdot \mathbf{V}. \quad (3.1.2)$$

Then, the dimension of the right-hand part of Equation 3.1.2 can be reduced to $r$ according to the Eckart-Young theorem [18]:
$$\mathbf{\Phi} \approx \mathbf{\Phi}^* = \arg\min_{\widetilde{\mathbf{\Phi}}} \left\| \mathbf{\Phi} - \widetilde{\mathbf{\Phi}} \right\|_F \quad (3.1.3)$$
$$s.t.\ rank(\widetilde{\mathbf{\Phi}}) = r,$$
where $\|\cdot\|_F$ is the Frobenius norm. In this manner, PCA can reduce the dimension of the high-dimensional system by using orthogonal projection. In the present work, rank $r$ indicates the number of observations in the spatiotemporal field. Hence, information in the corresponding $r$ rows of $\mathbf{\Psi}$, $\mathbf{S}$, and $\mathbf{V}$ will be extracted as $\mathbf{\Psi}_r$, $\mathbf{S}_r$, and $\mathbf{V}_r$, respectively. Also, the signal can be reconstructed as:
$$\mathbf{\Phi}^* = \mathbf{\Psi}_r \cdot \mathbf{S}_r \cdot \mathbf{V}_r \quad (3.1.4)$$

Then, a canonical measurement matrix $\mathbf{C}$ is used to select critical sampling locations sparsely, and the corresponding observation $\mathbf{y}$ can be simplified as:
$$\mathbf{y} = \mathbf{C}\phi = [\phi_{\gamma_1}, \phi_{\gamma_2}, \cdots, \phi_{\gamma_r}]^T, \quad (3.1.5)$$

where $\gamma = \{\gamma_1, \gamma_2, \cdots, \gamma_r\} \subset \{1, 2, \cdots, m\}$ is the set of indices for the selected sensors. In this manner, Equation 2.2.3 can be used to derive sampling locations from $\boldsymbol{\phi}$ efficiently.

Next, the observations in the field, $\phi_j \in \boldsymbol{\phi}$, can be represented as a linear combination of the basis and the coefficients, according to Equation 3.1.1 and Equation 3.1.3:

$$\phi_j = \sum_{k=1}^{r} \boldsymbol{\Psi}_{jk} \boldsymbol{a}_k \quad (3.1.6)$$

Observations in the sensor deployment locations can then be expressed as the linear combination of the canonical basis and the states according to Equation 3.1.5 and Equation 3.1.6:

$$y_i = \sum_{j=1}^{n} \mathbf{C}_{ij} \phi_j = \sum_{j=1}^{n} \mathbf{C}_{ij} \cdot \sum_{k=1}^{r} \boldsymbol{\Psi}_{jk} \boldsymbol{a}_k . \quad (3.1.7)$$

Equation 3.1.7 can be simplified as:

$$\boldsymbol{y} = \mathbf{C} \cdot \boldsymbol{\Psi}_r \cdot \boldsymbol{a} \quad (3.1.8)$$

According to Equation 3.1.1, $\boldsymbol{\phi}$ can be reconstructed by the basis coefficients $\widehat{\boldsymbol{a}}$:

$$\widehat{\boldsymbol{\phi}} = \boldsymbol{\Psi}_r \widehat{\boldsymbol{a}}. \quad (3.1.9)$$

Usually, only $r$ sensors are deployed in the field, and only the corresponding readings are obtained, which are denoted as $\mathbf{y} \in \mathbb{R}^r$. Hence, according to Equation 3.1.8 and Equation 3.1.9:

$$\widehat{\boldsymbol{\phi}} = \boldsymbol{\Psi}_r \widehat{\boldsymbol{a}} = \boldsymbol{\Psi}_r (\mathbf{C}\boldsymbol{\Psi}_r)^{-1} \mathbf{y}, \quad (3.1.10)$$

where $\boldsymbol{\Psi}_r \in \mathbb{R}^{n \times r}$ can be learned through SVD. Then, given $\boldsymbol{\Psi}_r$ and $\mathbf{y}$, the reconstruction performance is dependent on measurement matrix $\mathbf{C}$ ($\mathbf{C} \in \mathbb{R}^{r \times m}$). $\mathbf{C}$ can be optimized by finding the most informative locations in the field, such as the sensor deployment locations developed in the following section. Thus, the environment $\widehat{\boldsymbol{\phi}}$ can be reconstructed given $r$ observations $\mathbf{y}$, where $m \gg r$.

### 3.2 Data-driven Sparse Sensor Deployment with Informative Sampling

Optimal sensor deployment locations can guarantee the most accurate reconstruction of $\widehat{\boldsymbol{\phi}}$. For this reason, the locations of the sensor nodes should be optimized to acquire the most information from point measurements in sparsely selected sampling locations. Observations from the sparse sampling locations can then be used to reconstruct high-dimensional states, given the tailored SVD basis. Recall that $\gamma$ represents the structure of the canonical matrix $\mathbf{C}$, which gives the sampling locations in the field. Therefore, $\gamma$ affects the reconstruction performance. Thus, the optimal $\gamma$ and its corresponding sampling locations should provide the most information about the field.

Choosing a suitable number of sensors in the field while maintaining a low noise level in the data, is a challenge. Gavish et al. proved an optimal threshold, singular value hard thresholding (SVHT), that guarantees minimal signal reconstruction error [19]. While ensuring least reconstruction error, SVHT finds the optimal size of the tailored SVD basis. However, SVHT cannot find the optimal sensor deployment locations, since it compresses the original signal into a latent low-dimensional space, which cannot be sampled by sensors. Hence, the result provided by SVHT is regarded as the reconstruction target. In this manner, the number of sampling locations can be calculated, and $r$ rows of $\boldsymbol{\Psi}$, which correspond to the sensor deployment locations, should be obtained with maximum information in the spatiotemporal field.

Given the optimal number of informative sampling locations, it is essential to optimize the deployment locations. In the present paper, a simulated-annealing-based greedy approach is proposed to optimize the sampling locations. The sampling locations $\gamma$ can be optimized given the training dataset ($\boldsymbol{\Phi}_{train}$).

In Algorithm 1, line 1 initializes the algorithm by pre-processing the training data, where the training data can be obtained from the environmental model. The mean normalization is carried out to facilitate faster learning. Line 2 learns the training data by decomposing it into orthonormal matrices using SVD. Then, $r$ random sampling locations are selected from set $\mathcal{L}$ from line 3 to line 7, and they are transformed into the measurement matrix $\mathbf{C}_\gamma$ in line 8. Next, the spatiotemporal field is reconstructed according to Equation 3.1.10 in line 9 and line 10, and the reconstruction error ($\epsilon$) is calculated in line 11. Last, the algorithm iterates $T$ times to optimize the sampling locations from line 12 to line 22.

| Algorithm 1: Sampling location optimization |
|---|
| **Input**: $\boldsymbol{\Phi}_{train}$, $\mathcal{L}$, $r$, $m$, $T$ |
| **Output**: $\gamma$ |
| 1: Init.; |
| 2: $\boldsymbol{\Psi}, \mathbf{S}, \mathbf{V} = svd(\boldsymbol{\Phi}_{train})$; |
| 3: $\gamma = \emptyset$; |
| 4: **For** $i = 1, 2, 3, \cdots, r$ **do** |
| 5: $\quad \gamma_i \leftarrow rand_{m \in \mathcal{L}}(m)$; |
| 6: $\quad \gamma = [\gamma, \gamma_i]$; |
| 7: **End** |
| 8: $\mathbf{C}_\gamma \leftarrow canonical(\gamma)$; |
| 9: $\boldsymbol{\Phi}_{train}^{\gamma} \leftarrow \mathbf{C}_\gamma \boldsymbol{\Phi}_{train}$; |

10:  $\Phi^* \leftarrow \Psi_r(C_\gamma \Psi_r)^{-1} \Phi_{train}^\gamma$;
11:  $\epsilon = \text{MSE}(\Phi^*, \Phi)$;
12:  **For** $i = 1$ **to** $T$ **do**
13:   $\hat{\gamma} \leftarrow$ replace $\gamma_* \in \gamma$ with $l \in \mathcal{L}\backslash\gamma$;
14:   $C_{\hat{\gamma}} \leftarrow \text{canonical}(\hat{\gamma})$;
15:   $\Phi_{train}^{\hat{\gamma}} \leftarrow C_{\hat{\gamma}} \Phi_{train}$;
16:   $\Phi^* \leftarrow \Psi_r(C_{\hat{\gamma}} \Psi_r)^{-1} \Phi_{train}^{\hat{\gamma}}$;
17:   $\hat{\epsilon} = \text{MSE}(\Phi^*, \Phi)$;
18:   **If** $\hat{\epsilon} < \epsilon$ and $rand(0,1) \leq \rho$ **then**
19:    $\gamma = \hat{\gamma}$;
20:    $\epsilon = \hat{\epsilon}$;
21:   **end**
22:  **end**

During the optimization, the algorithm will first replace a random sampling location ($\gamma_*$) with a random location in set $\mathcal{L} \setminus \gamma$. By doing so, a new sampling location is tested for obtaining a lower reconstruction error. Then, the new measurement matrix $C_{\hat{\gamma}}$ is calculated, and the corresponding observations $\Phi_{train}^{\hat{\gamma}}$ and the reconstruction error $\hat{\epsilon}$ are computed. Last, the performance of the new sampling locations ($\hat{\gamma}$) is evaluated for the optimization. Algorithm 1 accepts better results according to a probability of $\rho$ to avoid local minima [20]. This criterion partially accepts better sampling strategies to jump out of the local minima, which guarantees the convergence of the reconstruction. Hence, after $T$ times of optimization, the sampling locations can be optimized, and the reconstruction error will be minimized.

In this manner, $r$ sampling locations selected by Algorithm 1 can be converted into the canonical measurement matrix $C$ to reconstruct the field. Given the observations at the selected sampling locations, Algorithm 2 can reconstruct the spatiotemporal field.

**Algorithm 2:** Spatiotemporal Field reconstruction

**Input**: $\gamma, \Psi_r, Y$
**Output**: $\hat{\phi}$
1: Init.;
2: $C_\gamma \leftarrow \text{canonical}(\gamma)$;
3: $\Theta = C_\gamma \Psi_r$;
4: **Foreach** $y \in Y$ **do**
5:  $\hat{\phi}_y \leftarrow \Psi \Theta^{-1} y$
6: **end**

As shown in Algorithm 2, the input of the algorithm is the optimized sampling locations $\gamma$, the trained principal basis $\Psi_r$, and the observation matrix $Y$. Line 1 initializes the variables of the algorithm. Then, the optimized sampling locations are transformed into the canonical measurement matrix $C_\gamma$. Next, a basis ($\Theta$) used for reconstruction is calculated as the product of $C_\gamma$ and $\Psi_r$, in line 3. In this step, the most important entities of the principal basis are extracted for the purpose of the spatiotemporal field reconstruction. Last, every snapshot in the observation matrix $Y$ is processed in line 5.

In this way, the spatiotemporal field can be reconstructed by using the observations from the calculated sparse sampling locations, and the whole spatiotemporal field can be reconstructed and predicted using limited observations.

## 4 Simulation

This section presents the simulation results of the spatiotemporal field reconstruction, using a National Oceanic and Atmospheric Administration (NOAA) dataset. The proposed algorithm is compared with state-of-the-art informative sensor deployment methods, and the reconstruction performance is analyzed based on the mean square error (MSE) between the reconstructed field and the ground truth.

### 4.1 Experimental Setup

The NOAA dataset covers the southeast Americas Seas region, which includes the Gulf of Mexico and the Caribbean Sea [21]. The daily sea surface temperature (SST) in 2017 was extracted from this dataset to model the environment.

The performance target and the benchmark algorithms for informative sensor deployment are as follows:

- **SVHT**. SVHT finds optimal thresholds for the dimensions of the tailored SVD basis that guarantee minimal signal reconstruction error [19]. SVHT reduces only the size of the SVD basis, and the resulting tailored SVD basis is used to represent a higher-dimensional space. Note that SVHT cannot find the sampling locations because it encodes the original signal into a latent low-dimensional space, which cannot be sampled by sensors. Therefore, the result generated by SVHT is regarded as the performance target.
- **RRC**: Rapidly-exploring random cycles (RRC) is a model-based method that generates sampling

locations with the lowest estimation error along with a cycled path within the space-filling tree [13]. The cycled sampling locations can be used to perform periodic sampling using a single robot.

- **INFO**: Informativeness maximization is a model-based approach that finds sampling locations with maximum mutual information in the field [6]. It generates a near-optimal sampling set that maximizes the mutual information of the sampling data.
- **RRTPI**: RRTPI is a model-based RRT method that efficiently solves geodesic-based exploration [22]. Ten locations with the highest mutual information are selected in the spatiotemporal field, and the exploring tree has a higher probability of exploring areas with higher information gain and efficiently finding the sampling locations with the lowest estimation error.
- **DRLT** DRLT is a model-based near-optimal spatiotemporal field estimator [26]. It seeks a near-optimal sensor deployment that has both the lowest estimation error and the maximum information. It utilizes deep reinforcement learning to improve the efficiency of the field exploring tree.
- **Q-DEIM** Q-DEIM is a data-driven approach to optimize the sensor deployment locations for spatiotemporal field reconstruction [3]. Similar to the proposed methods, it uses SVD to learn information from the training data. Then, it utilizes QR factorization to extract top $r$ essential sampling locations in the field. The spatiotemporal field is reconstructed based on the observations obtained from sampling locations.

The proposed and benchmark algorithms were implemented using MATLAB and run on a PC with 4.0 GHz Quad-Core CPU and 16 G memory.

### 4.2 Simulation Results

Fig.1 presents one ground truth snapshot of the spatiotemporal field for testing.

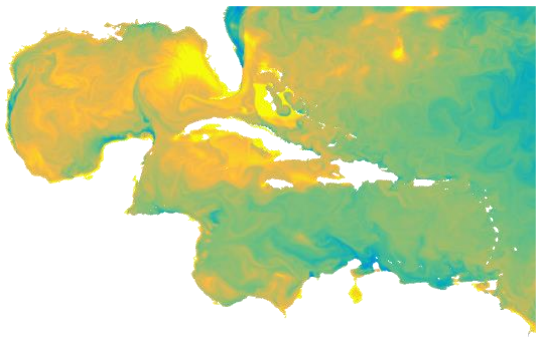

Fig.1 Ground truth.

Fig. 2 shows the reconstruction results generated by SVHT. This method has the best reconstruction results among all methods, as SVHT tailors only information that is less significant in the SVD basis. The input signal is compressed and reconstructed via an orthogonal projection with a tailored SVD basis. Therefore, a minimal amount of information is lost. However, although SVHT provides the best reconstruction results, it cannot be used for scheduling the sensor deployment locations. The information from the training dataset has been decomposed to a latent domain and cannot indicate the effectiveness of the sampling locations. Hence, the reconstruction of SVHT is regarded as the reconstruction target.

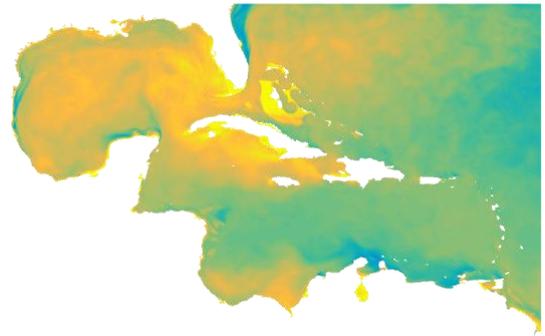

Fig.2 Reconstruction result generated by SVHT.

Fig. 3 and Fig. 4 present the reconstruction results provided by RRC and RRTPI. As shown in the figures, they collect only limited information and have the worst reconstruction performance among all methods. The poor performance is because RRC and RRTPI focus merely on minimizing the estimation error of the spatiotemporal field. Consequently, they do not collect enough information to represent the entire spatiotemporal field. Furthermore, RRC and RRTPI lead to a high reconstruction error, which are 39.37 and 64.17, respectively. As these two methods do not optimize the information gain in the field, sensor nodes may have been placed in locations that provide less effective observations.

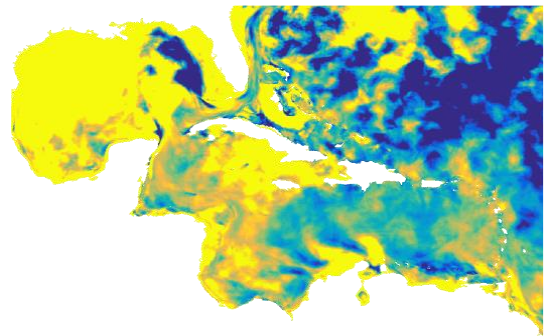

Fig.3 Reconstruction result generated by RRC.

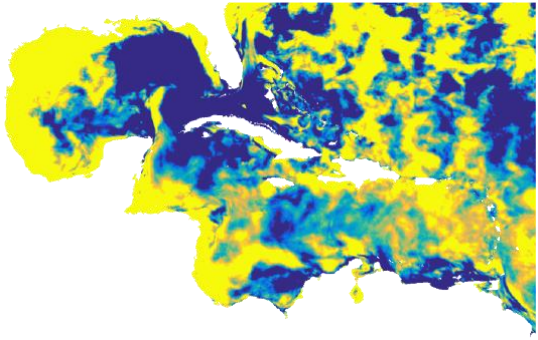

Fig.4 Reconstruction result generated by RRTPI.

Fig. 5 and Fig. 6 present the reconstruction results provided by INFO and DRLT. Both INFO and DRLT achieve significantly lower reconstruction error when compared to RRC and RRTPI, as they maximize the information gain in the field. The MSEs for the field reconstruction of INFO and DRLT are significantly reduced, which are 9.08 and 2.59, respectively. INFO focuses on maximizing the information gain from the field. As a result, INFO captures more information in the southern Caribbean Sea and achieves lower reconstruction error than RRC and RRTPI. However, the reconstruction error of INFO elevates in the northern Caribbean Sea and the Gulf of Mexico due to an inadequacy in the quality/quantity of information captured. In contrast, DRLT considers both estimation error and information gain to generate a more generalized sensor deployment strategy. As shown in Fig. 6, reconstruction in both southern and northern Caribbean Sea are superior to that provided by INFO.

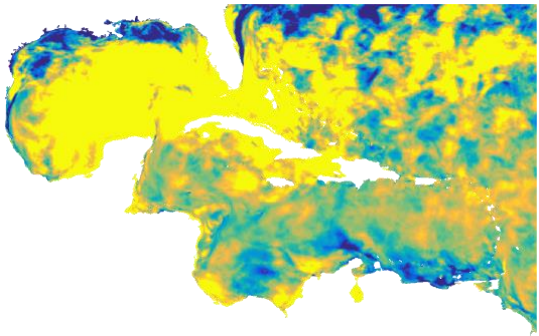

Fig.5 Reconstruction result generated by INFO.

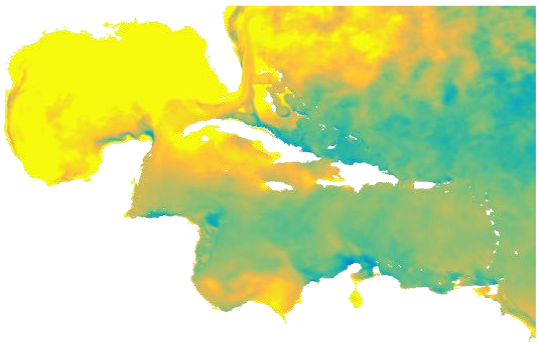

Fig.6 Reconstruction result generated by DRLT.

Fig. 7 and Fig. 8 present the reconstruction results generated by data-driven methods. It is clear that the data-driven approaches perform best in reconstructing a spatiotemporal field, which is also close to the performance target provided by SVHT. The reconstruction results provided by the data-driven approaches also indicate that these methods can restore complex environments, such as the SST over the Gulf of Mexico and the Northern Caribbean Sea. Moreover, compared to Q-DEIM, the proposed method achieves better reconstruction results in restoring complex environments. For example, the proposed method successfully reconstructs a sudden SST decrease near east Florida; in contrast, Q-DEIM does not perform well in this regard. Moreover, the proposed method significantly reduces the reconstruction error of the reconstructed snapshot. The MSEs for the field reconstruction of Q-DEIM and the proposed method are 0.4761 and 0.2071, respectively.

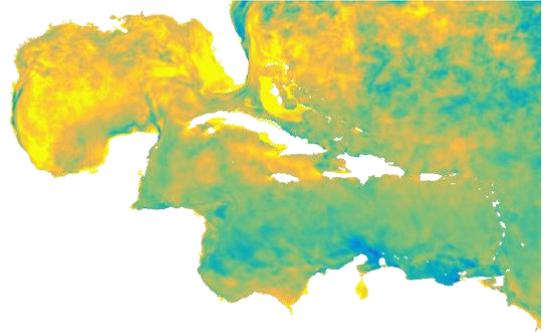

Fig.7 Reconstruction result generated by Q-DEIM.

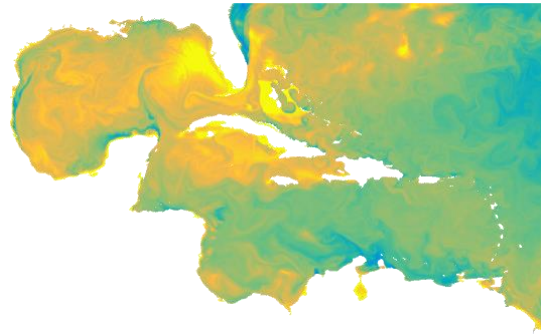

Fig.8 Reconstruction result generated by the proposed method.

Overall, the proposed method achieves the best performance in reconstructing a spatiotemporal field using limited sensor nodes.

Table 2 presents the MSE for the spatiotemporal field reconstruction using the proposed method and the benchmark algorithms. The proposed method outperforms both model-driven and data-driven benchmark algorithms by generating the most accurate spatiotemporal field reconstruction and achieving the lowest reconstruction error. As a comparison, the target reconstruction performance generated by SVHT is 0.12.

Table 2: MSE of the proposed method and the benchmark algorithms.

| Algorithms | MSE |
|---|---|
| RRC | 39.37 |
| RRTPI | 64.17 |
| INFO | 9.08 |
| RRLR | 2.59 |
| Q-DEIM | 0.4761 |
| Proposed method | 0.2071 |

Fig.9 presents the reconstruction error during the optimization of the sensor deployment locations. As indicated in the figure, the mean square error decreases monotonically as the number of iterations increases. Also, the reconstruction error converges after 1000 rounds of iteration.

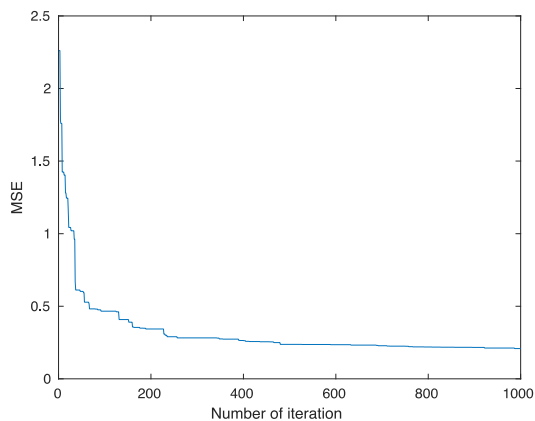

Fig.9 Reconstruction error during training.

# 5 Conclusion

This paper proposed a data-driven sensor deployment strategy for spatiotemporal field reconstruction. The proposed method selected the most informative sampling locations to represent the entire spatiotemporal field, and then reconstruct the spatiotemporal field according to the sampling data. Simulation results showed that the proposed strategy achieved the best spatiotemporal reconstruction performance among all methods considered in this study. Also, the reconstruction accuracy of the proposed method was high compared with the target performance generated by SVHT. Therefore, the reconstruction of a spatiotemporal field in the proposed method could be further improved. In general, model-based sampling methods may result in reconstruction inaccuracy when the environmental model cannot adequately express a highly dynamic field of interest; hence, data-driven approaches are more suitable for sparse sensor deployment.

## Author Biographies

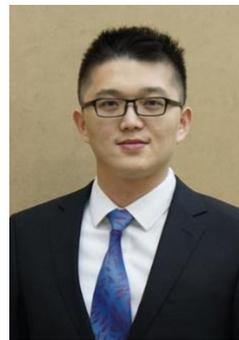

**Jiahong CHEN** received the B.E. degree in software engineering from Xiamen University, Xiamen, China, in 2015. He completed the Ph.D. degree in the Department of Mechanical Engineering, The University of British Columbia, Vancouver, Canada, and works as a post-doctoral researcher. His research interests include wireless sensor network, robotics, and machine learning.